\documentclass [12pt]{article}
\usepackage {longtable}

\begin{document}
Pacs 04.50. + h

\bigskip

\begin{center}
\textbf{To the problem of modeling of the gravitation and time.}
\end{center}

\begin{center}
P.Kondratenko\footnote{National Taras Shevchenko University of
Kyiv , Supreme Certifying Commission. 34, Khreshchatyk str., Kyiv,
01001, Ukraine. E-mail: pkondrat@unicyb.kiev.ua}
\end{center}

\bigskip

\begin{abstract}
Five-dimensional model of space-time (anti-de-Sitter space),
described within the dodecahedron symmetry is offered in the
article. According to the offered model graviton is the double
five-dimensional vortex (soliton) that provides its stability and
allows it to unite the Universe. The interaction between graviton
and particles of the Universe stipulates the discrete current of
time. The size of a quantum of time $\Delta \textit{t} \approx
10{}^{-103}$ seconds has been evaluated. Wave properties of the
mass particles follow from the model.

\end{abstract}

\bigskip

The history of physics progress in the 20-th century shows that
its rapid development always began with appearance of
revolutionary ideas (hypothesis). For example, it was so with
quantum physics. Similarly, the elementary particle physics,
investigating only external displays of particles, has received a
powerful impulse of the development with appearance of the
M.Gell-Mann and G.Zweig ideas on their internal constitution (idea
of quarks) [1].

The similar situation is observed in the theory of gravitational
field. There are a great number of theories concerning external
displays of a field. The requirements for development of
hypotheses touching models of gravitons and their physics have
ripened. The given investigation is just devoted to this problem.

While building the model, the author has paid attention to that
fact that main laws of the nature are the laws of symmetry, and at
the same time, our world, on the first view, appears to be
asymmetric.

First of all, it concerns the mass of free particle, energy and
temperature (effective temperature which can be assigned to the
system of excited particles in the state of the universe density
of population and which is not true temperature will not be
considered here). All of us have got used that a particle mass,
energy and temperature are always positive that disorders symmetry
of the Universe. Deficiency of symmetry also concerns the current
of time. We consider by default, that time is a characteristic of
space, which has a constant value of its vector. Nobody studied
physics of time.

It is important to introduce several postulates for the further
consideration:

1.   The modern theories of creation suppose that for the
description of natural phenomena it is necessary to bring in many
coordinates, from which only three are responsible for macroscopic
space, and the rest are enclosed in a ring with extremely small
radius. Since indicated three coordinates were created because of
a primary explosion, which led to creation of the Universe, it is
understandable, that before the explosion, these coordinates were
similar to others. After the explosion, which had a certain force,
they remained enclosed in a ring with big radius. So, \textbf{all
coordinates without exception are enclosed}.

2.   Numerous historical and geological data pointed out in the
literature (beginning from Plato and Pyphagoras up to the modern
theories of civilizations origin and of lithologic plates) testify
that the Earth is in the field, which is characterised by the
symmetry of a dodecahedron (regular icosahedron). So, \textbf{the
symmetry of a dodecahedron (in a local approximation)
 should
characterise physics of the Universe.}

3.   Common for us four-dimensional space corresponds to the
symmetry of a cube, which three edges at the apex correspond to
three spatial co-ordinates and quadrangular facet belongs to four
measurements. Accordingly, \textbf{the dodecahedron symmetry
testifies availability of three spatial co-ordinates} (three edges
at apex) \textbf{and five measurements} (pentagonal facet).

Having used the dodecahedron symmetry is, as indication of the
global symmetry, predetermining the law of the Universe, it is
possible with the help of a group theories to ascertain the types
and dimensions of subspace, responsible for time, mass,
gravitational waves and so on.

It was proved, that the group of a dodecahedron (Y) supposes the
existence of two one-dimensional ($\Gamma{}_{1g}$,
$\Gamma{}_{1u}$), four three-dimensional ($\Gamma{}_{2g}$,
$\Gamma{}_{2u}$, $\Gamma{}_{3g}$, $\Gamma{}_{3u}$), two four-
dimensional ($\Gamma{}_{4g}$, $\Gamma{}_{4u}$) and two five-
dimensional ($\Gamma{}_{5g}$, $\Gamma{}_{5u}$) representations
(subspaces). All indicated representations appear by pairs, one of
which is symmetric and the second one is asymmetric comparatively
with operation of inverse, which responds to the actual property
of the space. An attempt to consider these subspaces through the
symmetry of a cube (group $O{}_{h}$) has shown, that the
four-measuring subspace of group Y is parted thus on one and
three-dimensional spaces, and five-dimensional - on two and
three-dimensional spaces of group $O{}_{h}$. Such transformation
of subspaces leads to impossibility of the real processes
describing in $O{}_{h}$ group. These processes are described by
four-measuring and five-measuring representations of the groups Y
and being responsible for unity of the Universe.

\bigskip

Table 1. Rules of transformation of the representations in the
transition from group Y to group O${}_{h}$.

\newcommand{\PreserveBackslash}[1]{\let\temp=\\#1\let\\=\temp}
\let\PBS=\PreserveBackslash
\begin{longtable}
{|p{53pt}|p{53pt}|p{53pt}|p{86pt}|p{82pt}|} c & c & c & c & c
\kill \hline $\Gamma_{1g}$ $\rightarrow$ A${}_{1g}$& $\Gamma_{2g}$
$\rightarrow$ T${}_{1g}$& $\Gamma_{3g}$ $\rightarrow$ T${}_{1g}$&
$\Gamma_{4g}$ $\rightarrow$ A${}_{1g}$ + T${}_{1g}$& $\Gamma_{5g}$
$\rightarrow$ E${}_{g}$ + T${}_{2g}$ \\ \hline $\Gamma_{1u}$
$\rightarrow$ A${}_{1u}$& $\Gamma_{2u}$ $\rightarrow$ T${}_{1u}$&
$\Gamma_{3u}$ $\rightarrow$ T${}_{1u}$& $\Gamma_{4u}$
$\rightarrow$ A${}_{1u}$ + T${}_{1u}$& $\Gamma_{5u}$ $\rightarrow$
E${}_{u}$ + T${}_{2u}$ \\ \hline
\end{longtable}

\bigskip

Quantity and symmetry of the representations of the group Y gives
the foundation to presume that there are four sorts of matter:
matter (mass \textit{m}), antimatter (antimass $\tilde {m}$),
minus - matter (minus-mass $\bar {m}$) and anti-minus-matter
(anti-minus-mass $\tilde {\bar {m}}$). Thus, the magnitudes
$\textit {m}$ and $\tilde {m}$ are positive, and $\bar {m}$ and
$\tilde {\bar {m}}$ is negative. This will provide a complete
symmetry of the Universe concerning mass. As negative energy of
free particles corresponds to the negative mass of these
particles, this ensures symmetry of the World concerning both the
energy and temperature. As follows from Table 1, the subspaces
$\Gamma{}_{2g,u}$ and $\Gamma{}_{3g,u}$ of the group Y correspond
to the same subspace $T{}_{1g,u}$ of group $O{}_{h}$ that hindered
to bring a pair of negative masses within the framework of the
four-dimensional space.

\textbf{The world must be integrated; otherwise, it has no right
to exist.} So, there is a parameter (field, interaction),
responsible for wholeness of the world. Such interaction must be
spread instantaneously in our usual time; otherwise the wholeness
is lost. For the spread description of such interaction the
additional temporary dimension it is offered to bring in (besides
usual time t we shall bring in orthogonal to it time coordinate
$\tau$). The gravitation field with its quantum - graviton is
logically considered a carrier. The introduction in consideration
of two time co-ordinates testifies that we have space de-Sitter II
of (anti-de-Sitter space [2]) kind. Naturally, such space, with
the availability of spherical symmetry is unlocked that
contradicts to the first postulate. The decrease in the space
symmetry this paper makes space enclosed again [3].

The symmetry of a graviton must correspond to the representation
of maximal dimensionality inas- much as the role of the World
combining is assigned to it. The provision of the interaction
demands that the graviton description must be performed by three
spatial co-ordinates and two times (\textit{t} and $\tau$). The
instantaneity (in time \textit{t}) of the interaction transfer
superimposes an additional demand to graviton: it must
 have zero
mass. So, it cannot be a source of the secondary gravitational
radiation. The requirement of zero mass for graviton may be
provided depicting the graviton as the two-particle soliton in the
structure of which there is the mass \textit{m} and minus-mass
$\bar {m}$, so, the sum of masses making up the soliton is equal
zero. As the graviton must have properties of a wave (in time
$\tau$), the graviton components should be figured of as a stable
pair of the vortexes of cyclone - anticyclone type. Such pair of
the vortexes always has finite excitation energy (unlike of a
single vortex, the energy of which in an equilibrium state is
infinite), just what stabilises it [4]. It is known, that an
anticyclone twisting in the surface of the Earth as the right
screw, success up the air, due to what the increased pressure is
always in its centre. Similarly, the cyclone creates airflow along
its axis downwards that results in pressure drop in its activity
region. So, the pair cyclone - anticyclone will be compulsorily
combined by the third vortex, which raises its stability and makes
it a multidimensional soliton.

We assume the material Universe consisting of three components
(\textit{m}, $\bar {m}$ and \textit{m}), disjointed by time
intervals $\Delta t/2$, where $\Delta t$ is quantum of time (state
$\textbf {A}$) to model of discrete current of time (quantum of
time). In such case, the complete mass equals $\textit {m}$. The
graviton, which is in the past as to the matter, interacts with
mass, which responds to time \textit{t}=0, adsorbs it (naturally
the vortex $\bar {m}$ is adsorbed, and the vortex ${m}$ blurs the
function of element mass ${m}$ in time). The movement of the
graviton along closed time co-ordinate ${t}$ ensures its total
absorption of mass ${m}$. This adsorption transfers the system in
state $\textbf {B}$, in which the first element ${m}$, blurred in
time, is overlapped with the second device $\bar {m}$. The
structure instability arises, in which this pair (${m,} \bar {m}$)
disappears, and instead of it a new pair ($\bar {m}, {m}$) with
time co-ordinates 3$\Delta t/2$ ($\bar {m}$) and 2$\Delta t$
(${m}$) (state  $\textbf {C}=\textbf {A}+\Delta t$), appears
symmetrically relatively to the third element ${m}$ as well as a
new graviton, displaced relatively to the first graviton in time
by $\Delta t$. The process will repeat infinitely.

For current of time \textit{t} in an opposite direction the mass
must have structure ($\bar {m} ,m, \bar {m}$). In such case (state
$\textbf {B}$), the graviton must be above upper element of the
mass $\bar {m}$ (that is, hereafter from the point of view of
substance). Then this element will adsorb a vortex m of a
graviton. Further process will run as above described, with
formation of a new state, displaced by -$\Delta t$, that is, for
negative masses the time from the point of view of positive masses
will move to the past. Similarly, from the viewpoint of positive
masses, for negative masses the processes of the radiation will be
change by the processes of adsorption and vice verse, as the signs
of energies connected with matter, during the transition to minus
- material will be changed. The special role is assigned to a
quantum, which from the viewpoint of the world of substance and
the world of minus-material moves in opposite directions of time
and space that ensures its identical perception from both points
of reference.

The quantity of time quantum $\Delta t$ can be estimated, starting
from the formula $\Delta t = h/M_{U}\cdot c^{2}$, where $M_{U}$ is
mass of the Universe. Here, the supposition is made that Plank
constant is also attributed to discreteness of time. Considering
it in a rough approximation the Universe to be spherical (such
approximation contradicts to the first postulate, mentioned above)
with radius $R_U \approx 10^{10}$ light year = $10^{28}$ cm, and
average density of a matter equal to critical $\rho _m = \rho_c =
2\cdot 10^{-29}$ g/cm${}^{3}$, we fined $M_U = 4\pi \rho_c R_U^3
/3 = 10^{53}$ kg [5]. Hence, $\Delta t \approx 10^{-103}$ seconds.
This quantity really will be quantum of time combining and
synchronising the Universe.

It is known, that from a gravitational constant G, the speed of
light \textit{c} and the Plank constant \textit{h} it is possible
to form the length  $l_g = \sqrt{Gh/c^3} = 1,6 \cdot 10^{-33}$ cm,
which is named ``fundamental'' and ``gravitational'' [6] in
literature. Time $t_g = l_g/c = 5\cdot 10^{-44}$
 s, the value of which exceeds
by 60 orders the value of time quantum corresponds to this length.
So, ``gravitational length'' does not relate to the structure of
time-space.

At the same time, each material particle of the Universe will have
its time interval $\Delta t_i = h/m_i c^2$, responsible for wave
properties of elementary particles. For an electron ($m_e = 9,1085
\cdot 10^{-31}$ kg) the value $\Delta t_e = 0,809 \cdot 10^{-20}$
s, that by 83 orders exceeds the value of the time quantum. Let's
note, that the value $\Delta t_e$ on its origin has nothing common
with the period of the de-Broyl wave, though it will be close to
its relativistic velocities. It is understandable, that for the
synchronisation of the Universe it is necessary that $\Delta t_i =
N_i \Delta t$ where $N_i = M_U /m_i$ should be whole number. From
the latter ratio follows that the ratio of two particle mass $m_i
/m_j = N_j/N_i$ is rational number. This also will concern the
identical particles moving with different velocities, that is, the
travelling particle speed can vary only discretely. It is truth,
the discreteness step will be miserable and unnoticeable, that
ensures quasi-continuous dependence of the particle mass on its
motion velocity.

If the unity of the world could be absent then the graviton
radiated by an elementary particle would interact only with the
same particle, inasmuch as other elementary particles could exist
in other time points. This could result in the absence of the
gravitation interaction and, as the consequence, in the
disappearance of the material world. So, the unity is absolutely
necessary and it is ensured in the whole Universal by the mutual
sensation of all identical elementary particles. And this, in its
turn, will lead to that each elementary particle with a particular
phase of its existence function will be represented in every
moment of the discrete time (e.g., the function of the particle
existence may be described by the expression $\psi_i = a \cdot
exp(- \textit{i} \omega_i t$), where $\omega_i = 2\pi/ \Delta
t_i$, $a = c\cdot \sqrt {m_i /h}$ - normalised factor).

Thus, the used symmetry of a dodecahedron as local symmetry of the
Universe, in this paper, has allowed to bring to the symmetry of
the structure of a substance, scale of energy and time as well as
to proposal of the graviton model of providing unity of the
Universe and discrete current of time. The model becomes the basis
of wave properties of elementary particles, that is, expostulates
of their wave nature.

Inasmuch as for the creation of the gravitation model and time the
dodecahedron symmetry of the Universe field is used in which the
Earth exists the opposite conclusion corresponding to the reality:
the dodecahedron symmetry of the Earth surface must follows from
the proposed model.

\newpage
\textbf{References.}

\bigskip

[1]. F.J.Yndurain. Quantum Chromodynamics. An Introduction to the Theory of
Quarks and Gluons. Springer-Verlag. New York - Berlin - Heidelberg - Tokyo.
1983, 288 pp.

[2]. S.W.Hawking, G.F.R.Ellis. The Large Scale Structure of Space-Time.
Cambridge Univ. Press, 1973. 431 pp.

[3]. L\"{o}bell F. Beispiele geschlossener drei-dimensionaler
Clifford-Kleinsche R\"{a}ume negativer Kr\"{u}mmung. / Ber.
Verhandl. SÄchs. Akad. Wiss. Leipzig, Math., Phys. K1., 1931.
B.83, S.167-174.

[4]. V.V.Meleshko, M.Yu.Konstantinov. Dynamics of vortex
structures. Kyiv: <Naukova dumka>, 1993, 278 pp.

[5]. The tables of physical quantities. The manual by I.K.Kikoin. Moscow:
Atomizdat, 1976, 1006 pp.

[6]. V.L.Ginzburg. About physics and astrophysics. Moscow: ``Nauka'', 1980,
156 pp.

\end{document}